\documentclass[prb,superscriptaddress,twocolumn,amsfonts,amsmath,a4paper]{revtex4}
\usepackage{graphicx}

\begin{document}

\title{Logarithmic Relaxation in a Kinetically Constrained Model}

%
%
\author{Angel J. Moreno}
\email[Corresponding author: ]{wabmosea@sq.ehu}
\affiliation{Donostia International Physics Center, Paseo Manuel de Lardizabal 4,
20018 San Sebasti\'{a}n, Spain.}
\author{Juan Colmenero}
\affiliation{Donostia International Physics Center, Paseo Manuel de Lardizabal 4,
20018 San Sebasti\'{a}n, Spain.}
\affiliation{Dpto. de F\'{\i}sica de Materiales, Universidad del Pa\'{\i}s Vasco (UPV/EHU),
Apdo. 1072, 20080 San Sebasti\'{a}n, Spain.}
\affiliation{Unidad de F\'{\i}sica de Materiales, Centro Mixto CSIC-UPV, 
Apdo. 1072, 20080 San Sebasti\'{a}n, Spain.}

\date{\today}
\maketitle

Very recently, we have reported a numerical investigation on a simple bead-spring model for polymer blends
with components of very different relaxation times \cite{blendpaper}. 
At moderate and high dilution of the fast component, the slow component acts as a confining medium
for the former. In this situation, dynamic correlators for the fast component display non standard
relaxation features. Instead of the usual two-step decay observed for dense colloidal systems
or deeply supercooled liquids, dynamic correlators do not exhibit a defined plateau and the long-time decay
is highly stretched. Decreasing temperature yields a striking concave-to-convex crossover 
in the shape of the correlators. At some intermediate temperature, the decay is purely logarithmic over
3-4 time decades. Atomistic simulations on real polymer blends show a qualitatively similar behavior \cite{genix}.
These results exhibit striking analogies with predictions of the Mode Coupling Theory (MCT)
for state points close to a higher-order transition \cite{schematic,sperl}.
In particular, asymptotic laws predicted by this MCT scenario for dynamic correlators
provide an excellent description of simulation results for the fast component \cite{blendpaper}. 
 
Higher-order MCT transitions have been strictly
derived for simple models of short-ranged attractive colloids \cite{sperl,fabbian,bergenholtz,dawson}.
For these systems, such transitions arise in regions of the control parameter space where 
reversible bond formation (originating from the short-ranged attraction) and hard-sphere
repulsion compete as different mechanisms for dynamic arrest. 
The observed strong dynamic analogies with
polymer blends \cite{blendpaper} suggest that this anomalous
relaxation scenario might be a general
feature of systems where dynamic arrest is driven by several competing mechanisms 
of different characteristic lenghts. For the fast component in polymer blends
we have suggested confinement and bulk-like caging \cite{blendpaper}, respectively induced
by the host matrix formed by the slow component, and by neighboring particles of the fast component.

Kinetically constrained models \cite{fredrickson,sollich} are used as coarse-grained
pictures for relaxation in supercooled liquids.
In these models, the liquid structure is substituted by an array of cells. The cell size roughly corresponds
to a density correlation length. 
A spin variable is assigned to each cell, with values 0 or 1 denoting
respectively unexcited and excited local states in a mobility field. Change in local mobility
(spin flip) for a given cell is permitted according to kinetic constraints 
determined by the mobilities of neighboring cells. Propagation of mobility, leading to structural relaxation,
occurs via dynamic facilitation: microscopic regions become temporarily mobile only if neighboring regions are mobile.
Kinetically constrained models provide in a simple way an important feature of glass-forming 
liquids: the increasing 
of dynamical correlation lengths with decreasing temperature,
leading to dynamic heterogeneity \cite{prnat,prlfac,prefac,nef}.

\begin{figure}
\includegraphics[width=0.81\linewidth]{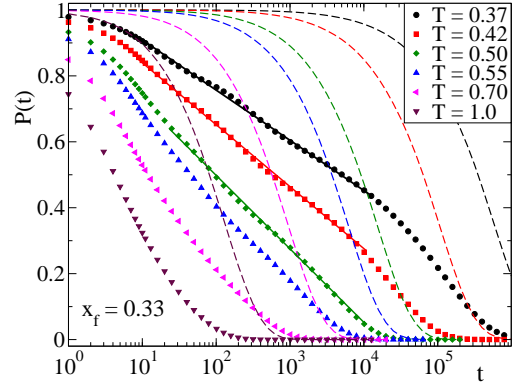}
\newline
\caption{Temperature dependence of the mean persistence function $P(t)$ at composition
$x_{\rm f}=0.33$. Symbols correspond to fast cells. Dashed lines correspond to slow cells
(same temperatures as for the former ones, decreasing from left to right).
Solid straight lines indicate nearly logarithmic relaxation.}
\label{fig1}
\end{figure}

Some of the anomalous dynamic features displayed by short-ranged attractive colloids,
as logarithmic relaxation or reentrant behavior of the diffusivity, have been recently 
reproduced by a simple model based on this latter approach \cite{geissler}.
Motivated by this fact and by the mentioned dynamic analogies between short-ranged attractive
colloids and polymer blends, in this Note we investigate a simple kinetically constrained model
aimed to reproduce dynamic features in real binary systems with components of very different relaxation rates.
Results reported here exhibit strong analogies with dynamic correlators in polymer blends. 

We have carried out Monte Carlo (MC) simulations on a variation (see below) 
of the north-or-east-or-front (NEF) model
recently investigated by Berthier and Garrahan \cite{nef}. In these three-dimensional model, spin flip 
of a given cell is only permitted if there is at least one excited neighboring cell
in the north-or-east-or-front direction \cite{fragile}. More specifically, if the cell is denoted
by its position vector ${\bf i}=$ ($i_{\rm x}$,$i_{\rm y}$,$i_{\rm z}$),
spin flip is permitted according to the following rules:

i) At least one of the neighboring cells in the north ($i_{\rm x}$,$i_{\rm y}$,$i_{\rm z}+1$),
or east ($i_{\rm x}$,$i_{\rm y}+1$,$i_{\rm z}$),
or front ($i_{\rm x}+1$,$i_{\rm y}$,$i_{\rm z}$) direction is excited, i.e., it has spin 1.

ii) If i) is fulfilled, spin flip is accepted according to the Metropolis rule \cite{frenkel}.
Hence, $1 \rightarrow 0$ is always accepted, while $0 \rightarrow 1$ is accepted with a 
thermal probability $[1+\exp(1/T)]^{-1}$, where $T$ is the temperature. 

In the present work we investigate a binary mixture of cells with the same population
of excitations, $[1+\exp(1/T)]^{-1}$, but different rates for excitation,
$\exp(-E/T)[1+\exp(1/T)]^{-1}$, and de-excitation, $\exp(-E/T)$, of cell mobility.
This choice fulfills detailed balance. We use $E =$ 0 and 3 for the different components, 
which are respectively denoted as ``fast'' and ``slow'' cells. Hence, spin flip rules for 
the fast component are the same as in the original NEF model, aimed to reproduce bulk-like dynamics.
The probability of spin flip for slow cells is decreased by a factor $\exp(-3/T)$ as compared
to that of fast cells, providing a large separation in the time scales of both components (see below).  

The mixture composition, $x_{\rm f}$, is defined as the fraction of fast cells.
We investigate a wide range of composition and temperature as control parameters.
A square box of side $N = 50$ cells is used. Periodic boundary conditions
are implemented. Slow and fast cells are randomly distributed according to the selected value of $x_{\rm f}$.
Time is given in MC steps.
Within each MC step a total of $N^3$ trials (one for each cell) are performed.

\begin{figure}
\includegraphics[width=0.81\linewidth]{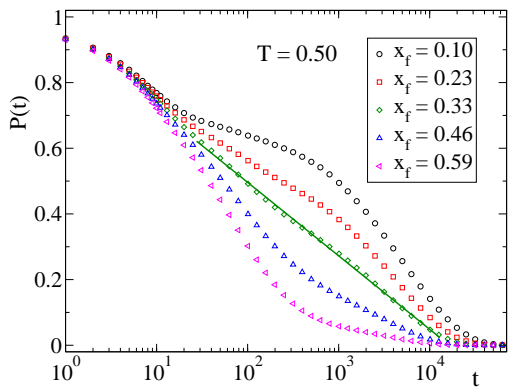}
\newline
\caption{Symbols: composition dependence of  $P(t)$, at $T=0.50$, for the fast cells.
The straight line indicates nearly logarithmic relaxation.}
\label{fig2}
\end{figure}

We have evaluated the mean persistence function $P(t) = \sum_{{\bf i}} p({\bf i} ; t)$, where  $p({\bf i} ; t)$
is the persistence function of the cell ${\bf i}$ at time $t$. It takes the value 1 if no
spin flip has occurred for that cell in the interval [0,$t$]. If one or more spin flips have occurred,
it takes the value 0.
Fig. \ref{fig1} shows results for $P(t)$ at composition $x_{\rm f} = 0.33$ and different temperatures.
Note that the first decay usually observed for any correlator in real systems 
---corresponding to the transient regime at microscopic times--- is obviously
absent due to the coarse-grained character of the model.
The introduction of very different rates for cell spin flip provides very different relaxation times for
fast and slow cells. Slow cells display standard relaxation, as observed for the slow component
in polymer blends \cite{blendpaper}.  However, fast cells display rather different relaxation features. 
In complete analogy with results for the fast component
in polymer blends \cite{blendpaper,genix}, $P(t)$ exhibits a concave-to-convex crossover with decreasing temperature.
At an intermediate temperature, pure logarithmic relaxation is observed over a time interval of near three decades.
This behavior is also observed by fixing temperature and varying the mixture composition (see Fig. \ref{fig2}).
It is worthy of remark that the features displayed in Figs. \ref{fig1} and \ref{fig2} are not specific of $P(t)$.
Qualitatively analogous results are exhibited by other dynamic correlators, as persistence correlators
for distinct cells, or intermediate scattering functions \cite{chandler}.  

In summary, the highly non-trivial anomalous relaxation features recently reported for the fast component
in polymer blends can be qualitatively reproduced by a simple kinetically constrained model.
The fact that another recent work within this approach \cite{geissler} also reproduces unusual relaxation
features for short-ranged attractive colloids suggest the relevance of the dynamic facilitation picture
for systems with several competing mechanisms for dynamic arrest. 

We thank J. P. Garrahan and L. Berthier for useful discussions.



\begin{thebibliography}{9}
\bibitem{fragile} Directionality in the kinetic constraints is imposed
to mimic fragile liquid behavior \cite{nef}.
\end{thebibliography}
\end{document}